\input{aipcheck}

\documentclass[
    ,final            
  ]
  {aipproc}

\layoutstyle{8x11double}

\begin{document}

\title{Center vortex model and the G(2) gauge group}

\classification{11.15.Ha; 12.38.Aw; 12.39.Pn; 12.38.Gc}
\keywords      {QCD, phenomenological model, center vortex, confinement}

\author{Sedigheh Deldar}{
  address={Department of Physics, University of Tehran, P.O. Box 14395/547, Tehran 1439955961,
Iran}
}

\author{Hadi Lookzadeh}{}

\author{Seyed Mohsen Hosseini Nejad}{}

\begin{abstract}
 The thick center vortex model is applied to G(2) gauge group to obtain the potentials between static sources of the fundamental and adjoint representations. The group G(2) has only one trivial center element and therefore it does not have any vortices which are defined based on non trivial center elements. To obtain the potential from the thick center vortex model, the idea of the vacuum domain structure is used. The intermediate string tensions from this model are in rough agreement with the G(2) lattice results and the Casimir ratio. We argue that the SU(3) subgroup of G(2) may be responsible for the linear potential at the intermediate distances.
\end{abstract}

\maketitle


\section{Introduction}

The thick center vortex model is a phenomenological model which gives a correct behavior for the potentials between static sources at intermediate and large distances. Confinement results from random fluctuations in the number of center vortices linked to the Wilson loops. 
The magnetic flux of a vortex is quantized in terms of the center elements of the gauge group. Therefore, confinement is not expected for a group without any non trivial center element. G($2$) is a group which has only a trivial center element. It seems interesting to investigate how the thick center vortex model works for this exceptional group. It has already been observed in lattice calculations \cite{olejnik2008} that for this group, the potential between a pair of quark antiquark has a linear regime and as a result the quarks are confined. The interesting question may arise: how can one get the confinement without the center elements?

In this paper, we show that using modified thick center vortex model with the idea of vacuum domain structures instead of vortices, one can produce potentials in rough agreement with lattice results. We discuss about the possibility that the confinement in the group G($2$) arises as a result of decomposition into its SU($3$) subgroup.

\section{The group G($2$)}

G($2$) is one of the simplest exceptional Lie groups which like SU($3$) has rank $2$. It is the simplest in the sense that its universal covering group is the group itself. It has only trivial center element. This group has $14$ generators and thus $14$ objects in the adjoint representation. The dimension of the fundamental representation is $7$. The group is real and is a subgroup of SO($7$) of rank $3$ with 21 generators. The determinant of the $7 \times 7$ real orthogonal matrices $\Omega$ of the group $SO(7)$ is 1 and:
\begin{equation}
\Omega_{ab} \Omega_{ac} = \delta_{bc}.
\end{equation}
The $G(2)$ subgroup elements satisfy a constraint called the 
cubic constraint:
\begin{equation}
T_{abc} = T_{def} \Omega_{da} \Omega_{eb} \Omega_{fc}.
\end{equation}
$T$ is a totally anti-symmetric tensor and its non-zero elements are:
\begin{equation}
T_{127} = T_{154} = T_{163} = T_{235} = T_{264} = T_{374} = T_{576} = 1.
\end{equation} 
The last two equations lead to the reduction of the 21 generators of SO($7$) to the $14$ for G($2$) gauge group.

To apply G($2$) to the thick center vortex model, we need the Cartan subalgebra of the group. Since the rank of the group is $2$, only two of the generators can be diagonalized simultaneously. Because  SU($3$) is a subgroup of G($2$), we make those two generators out of the diagonalized SU($3$) Gell-Mann generators:
\begin{equation}
\label{cartan}
\Lambda_a = \frac{1}{\sqrt{2}} \left( \begin{array}{ccc} \lambda_a & 0 & 0 \\ 
0 & \; -\lambda_a^* & 0\\ 0 & 0 & 0 \
\end{array} \right).
\end{equation}
Here $\lambda_a$ ($a=3,8$) are the two diagonal  
$SU(3)$ generators. 

Under $SU(3)$ subgroup transformations, the $7$th and $14$th dimensional 
representations  of G($2$) decomposes into the SU($3$) fundamental and adjoint representations:
\begin{equation}
\label{7t03}
\{7\} = \{3\} \oplus \{\overline 3\} \oplus \{1\} ,
\end{equation}
\begin{equation}
\label{14to3}
\{14\} = \{8\} \oplus \{3\} \oplus \{\overline 3\}.
\end{equation}
The second equation may be interpreted as that the $14$ gluons of G($2$) consist of the usual $8$  gluons of SU($3$) plus $6$ additional gluons which transform like the SU($3$) fundamental quark and antiquark. One of the differences between the $6$ gluons and the SU($3$) quarks is that the former ones  are bosons while the latter ones are fermions. Quark and antiquark in the G($2$) group are the same. This is  because all G($2$) representations are real and thus the $\{7\}$ representation is equivalent to its complex conjugate.

\section{Vacuum structure and thick center vortex model}

The thick center vortex model is a phenomenological model which gives the potential between a pair of static quark antiquark in the fundamental and higher representations of SU($N$) gauge group. The potential is obtained based on the interaction between the Wilson loops and the topological field configurations of the vacuum named thick center vortices. The center-vortex model introduced in the late 1970's \cite{Hoof79} and it gave the potential between quarks in the fundamental representation. This model has been developed by Faber {\it {et al.}} \cite{Fabe98} to the thick-center-vortex model to obtain  linear potentials for higher representations. The vortex model states that the QCD vacuum is filled by  some special line-like (in three dimensions) or surface-like (in four dimensions) objects, which carry a magnetic flux quantized in terms of the center elements of the gauge group. 
The inter-quark potential induced by the thick vortices is obtained by:
\begin{equation}
\label{potential}
V(R) = \sum_{x}\ln\left\{ 1 - \sum^{N-1}_{n=1} f_{n}
(1 - {\mathrm {Re}} {\cal G}_{r} [\vec{\alpha}^n_{C}(x)])\right\}.
\end{equation}
$x$ is the location of the center of the vortex and $C$ indicates
the Wilson loop and ${\cal G}_{r}$ is defined as:
\begin{equation}
{\cal G}_{r}[\vec{\alpha}] = \frac{1}{d_{r}}
{\mathrm {Tr}} \exp[{\mathrm {i}}\vec{\alpha} . \vec{H}],
\label{gr1}
\end{equation}
$d_{r}$ is the dimension of the representation, $f_{n}$ is the
probability that any given unit is pierced by a vortex type $n$ and $\{H_{i},i=1,2,...,N-1\}$  are generators spanning the Cartan subalgebra.
The parameter $\alpha_{C}(x)$  describes the vortex flux distribution and
depends on the vortex location. The profile of the vortex should be chosen such that vortices which pierce the plane far outside the loop do not affect the loop.
On the other hand, if the vortex core is entirely contained within the loop, it will derive a  maximum multiplicative factor $\exp({\frac{2\pi i n}{N}})\in Z_{n}$  $(n=1,2,...,N-1)$ corresponding to the center elements. For the limit when the spatial size of the Wilson loop goes to zero, $\alpha$ should be zero, as well. There are many mathematical functions which satisfy these conditions. 

The magnetic flux is quantized in terms of the center elements of the gauge group. Therefore, the model can be applied to those groups which have non trivial center elements. However, in Ref. \cite{Greensite2007}, in order to increase the length of the linear part of the potential, the model has been modified by using both the trivial and non trivial center elements of the SU($2$) gauge group.  This means that one would have another type of the vortex with the probability $f_{0}$ in Eq. (\ref{potential}). It is called vauum domain in Ref. \cite{Greensite2007}. The non trivial center elements of the group which have been called vortices in the thick center vortex model, are called domains, as well. 

\begin{figure}
\label{longpot}
  \includegraphics[height=.23\textheight]{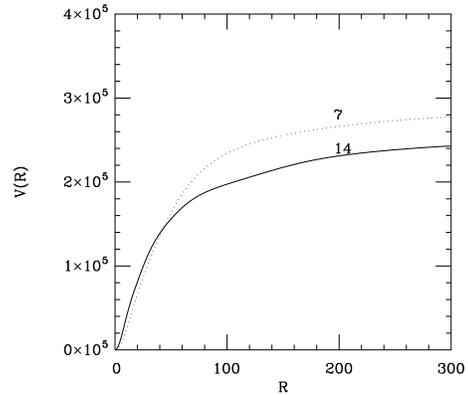}
  \caption{Potentials between two static sources of the $7$th and $14$th dimensional representations. $7$ is screened at a higher energy than the $14$. At intermediate distances a linear behavior is observed for both representations.}
\end{figure}

In this paper, we apply the idea of the domain structure to the thick center vortex model for G($2$) gauge group by rewriting Eq. (\ref{potential}) :
\begin{equation}
\label{G2potential}
V(R) = \sum_{x}\ln\left\{ 1 - f_{0}(1 - {\mathrm {Re}} {\cal G}_{r} [\vec{\alpha}^0_{C}(x)])\right\}.
\end{equation}
Where $f_{0}$ is the probability that any given unit is pierced by a vauum domain. ${\cal G}_{r}$ is
the same as Eq. (\ref{gr1}) and $d_{r}$ is the dimension of the representations of G($2$). 

We have used the fluctuating flux of the Ref. \cite{Deldar2010} to remove the concavity of the potential of the adjoint representation. At large distances where the vortex is contained completely inside the Wilson loop, ${\cal G}_{r} [\vec{\alpha}]$ is normalized to $I$ and $\alpha$ obtains its maximum value. Fig. \ref{longpot} shows the potential of the fundamental and the adjoint representations versus $R$, the distance between the sources. The potentials are flat at large distances. In fact, when the distance between the sources increases, pairs of gluons can pop of the vacuum and by combining with the initial sources make them screened. The interesting point is that $7$ is screened by three $14$ while $14$ is screened by one $14$. Therefore, the sources of the fundamental representation are screened at higher potentials than $14$'s as shown in the figure.
\begin{eqnarray}
\label{product7}
\{7\} \otimes \{14\} \otimes \{14\} \otimes \{14\}=\{1\} \oplus ...
\end{eqnarray}
\begin{eqnarray}
\label{product}
\{14\} \otimes \{14\} =\{1\} \oplus ...
\end{eqnarray}

\begin{figure}
\label{shortpot}
  \includegraphics[height=.215\textheight]{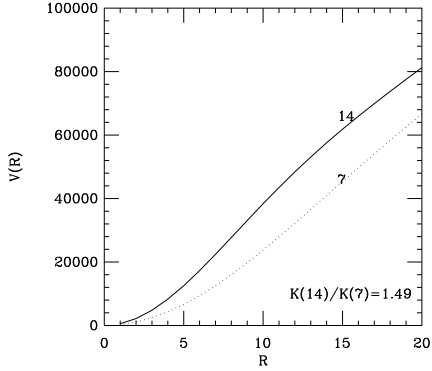}
  \caption{The ratio of the string tensions, $\frac{K_{14}}{K_\mathrm{7}}=1.49$, is in rough agreement with Casimir scaling which is 2.0.}
\end{figure}

Fig. \ref{shortpot} indicates the potential at small and intermediate distances. The ratio of the string tension of the adjoint representation to the fundamental representation is about $1.49$. It is in rough agreement with the ratio of the Casimir ratio of the two representation which is 2; and also with the ratio from the lattice \cite{olejnik2008} which is $1.88-1.96$.

To understand why G($2$) confines quark, we have calculated ${\mathrm {Re}} {\cal G}_{r} [\vec{\alpha}]$ for G($2$). It changes between $1$ and $-0.28$, where $1$ comes from the trivial center element and $-0.28$ can be obtained from the relation between the trace of the group G($2$) when it is decomposed into its SU($3$) subgroup, and the trace of the SU($3$) gauge group itself. In addition, at large distances where the vortex is completely inside the Wilson loop, we have normalized ${\cal G}_{r}[\vec{\alpha}]$ to the center elements of its SU($3$) subgroups instead of $I$:
\begin{equation}
\label{su3sub}
Z = \left( \begin{array}{ccc} zI_{3x3} & 0 & 0 \\ 
0 & \; -z^*I_{3x3} & 0\\ 0 & 0 & 0 \
\end{array} \right).
\end{equation}
Where $z=\exp \frac{2\pi i}{3} I$ and its complex conjugate are the non trivial center elements of SU($3$) gauge group. Figure \ref{7,7p} compares the potential of the $7$th dimensional representation of G($2$) with the potential obtained from the $7$ dimensional representation of its subgroup SU($3$) for which the flux is quantized to Eq. (\ref{su3sub}). For $25<R<35$, the slope of the potentials are equal. In other words the linear part of the potentials are parallel in this regime. This behavior is also observed for the adjoint ($14$) representation. One may argue that the G($2$) gauge group may be decomposed into its SU($3$) subgroup in this regime or in fact SU($3$) is dominant in this regime. Our results are also in agreement with the results by M. Pepe and  {\it {et al.}} \cite{Pepe2001} who have studied the confinement in G($2$) gauge group using both lattice gauge theory and the Higgs mechanism. 

\begin{figure}
  \includegraphics[height=.215\textheight]{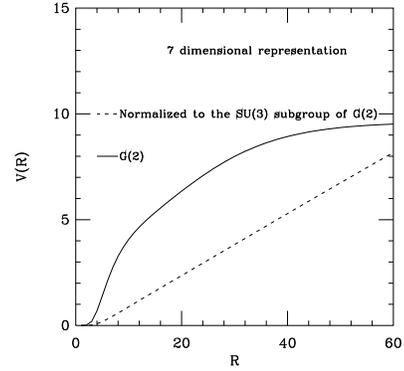}
  \caption{For $25<R<35$, the slopes of the potentials are equal, in other words the linear parts of the potentials are parallel in this regime.}
\label{7,7p}
\end{figure}

\section{Conclusion}

The thick center vortex model is applied to the G($2$) group. Even though this group dose not have any non trivial center element, but if one uses the domain structures instead of vortices, the model can be applied to the G($2$) gauge group, as well. We have obtained linear potentials for both fundamental and adjoint representations in rough agreement with the Casimir ratio and the lattice results. We interpret the linear regime as the regime where G($2$) is decomposed into its SU($3$) subgroup. As expected, the string tensions for both representations are zero at large distances. 


\begin{theacknowledgments}

We would like to thank M. Faber and S. Olejnik for the very useful discussions. We are grateful to the research council of University of Tehran for supporting this study.

\end{theacknowledgments}



\bibliographystyle{aipproc}   

\bibliography{sample}

\IfFileExists{\jobname.bbl}{}
 {\typeout{}
  \typeout{******************************************}
  \typeout{** Please run "bibtex \jobname" to optain}
  \typeout{** the bibliography and then re-run LaTeX}
  \typeout{** twice to fix the references!}
  \typeout{******************************************}
  \typeout{}
 }

\end{document}